# Use of Real-World Data and Real-World Evidence in Rare Disease Drug Development: A Statistical Perspective

Jie Chen[1], Susan Gruber[2], Hana Lee[2], Haitao Chu[4], Shiowjen Lee[3],
Haijun Tian[5], Yan Wang[3], Weili He[6], Thomas Jemielita[7], Yang Song[8],
Roy Tamura[9], Lu Tian[10], Yihua Zhao[11], Yong Chen[12],
Mark van der Laan[13], Lei Nie[3]

[1]ECR Global, Shanghai, China

[2]TL Revolution, Cambridge, MA, USA

[3]US Food and Drug Administration, Silver Spring, MD, USA

[4]Pfizer Inc, Groton, CT, USA

[5]Eli Lilly & Co., Indianapolis, IN, USA

[6]AbbVie, North Chicago, Illinois, USA,

[7]Merck & Co., Inc., Rahway, NJ, USA

[8]Vertex Pharmaceuticals, Boston, MA, USA

[9]University of South Florida, Tampa, FL, USA

[10]Stanford University, Stanford, CA, USA

[11]Flatiron Health, San Francisco, CA, USA

[12]University of Pennsylvania, Philadelphia, PA, USA

[13]University of California at Berkeley, Berkeley, CA, USA

*Correspondence to: Dr. Lei Nie, US Food and Drug Administration, Silver Spring, MD 20903. Email: lei.nie@fda.hhs.gov.

# Use of Real-World Data and Real-World Evidence in Rare Disease Drug Development: A Statistical Perspective

**Abstract**

Real-world data (RWD) and real-world evidence (RWE) have been increasingly used in medical product development and regulatory decision-making, especially for rare diseases. After outlining the challenges and possible strategies to address the challenges in rare disease drug development (see the accompanying paper), the Real-World Evidence (RWE) Scientific Working Group of the American Statistical Association Biopharmaceutical Section reviews the roles of RWD and RWE in clinical trials for drugs treating rare diseases. This paper summarizes relevant guidance documents and frameworks by selected regulatory agencies and the current practice on the use of RWD and RWE in natural history studies and the design, conduct, and analysis of rare disease clinical trials. A targeted learning roadmap for rare disease trials is described, followed by case studies on the use of RWD and RWE to support a natural history study and marketing applications in various settings.

**Key words**: Clinical trials, rare disease, real-world data, real-world evidence, targeted learning, causal inference.

## 1 Introduction

The 21st Century Cures Act (Cures Act), signed into law in December 2016, is designed to help accelerate medical product development and bring innovative medicines faster and more efficiently to patients who need them [1]. The Cures Act, building upon FDA's ongoing effort to incorporate the perspectives of patients into the development of medical products, authorizes the regulatory agency to modernize product development including use of real-world evidence (RWE) and to speed up the development and review of novel medical products. The FDA's RWE framework states that real-world data (RWD) can be used, among others, to assemble geographically distributed research cohorts in drug development for rare diseases and that the agency has accepted RWE to support drug product approvals for rare diseases [2]. In fact, the FDA has a history of using RWE to support drug development and regulatory decisions, especially for rare diseases; see, e.g., Jarow et al. [3], Feinberg et al. [4], Wu et al. [5], Gross [6], Liu et al. [7], Mahendraratnam et al. [8], Purpura et al. [9], Chen et al. [10], He et al. [11] and references therein.

Unlike clinical development of medical products for common diseases, developing drugs for rare diseases presents unique challenges. For example, small patient population sizes, lack of accurate diagnostic tools, limited understanding of natural history of the disease, lack of sufficient financial incentives to sponsors, etc. The Real-World Evidence (RWE) Scientific Working Group (SWG) of the American Statistical Association (ASA) Biopharmaceutical Sections (BIOP), under the auspice of the Public Private Partnership (PPP) of the Center for Drug Evaluation and Research (CDER) in the FDA, conducts a landscape assessment in rare disease drug development, provides a summary of challenges and possible strategies to address these challenges in rare disease drug development and regulatory decision-making [12]. In particular, they highlight, among others, the need on use of RWD and RWE along the development pathways to generate the totality of evidence for drugs treating rare disease. This research work by the ASA BIOP RWE SWG, as an accompanying paper of Chen et al. [12], is intended to provide more detailed discussion on the current landscape on the use of RWD and RWE in drug development and regulatory decision-making for rare diseases. A targeted-learning (TL)-based roadmap using RWD to generate RWE for rare disease drug development is described [13, 14], followed by case studies illustrating the use of RWD and RWE to support a natural history study and marketing applications in various settings.

The rest of this paper is organized as follows. Section 2 summarizes relevant regulatory guidance documents and frameworks by selected regulatory agencies on the use of RWD and RWE in drug development and regulatory decision-making. Section 3 describes current practice on the use of RWD and RWE in natural history studies and the design, conduct, and analysis of clinical trials for rare diseases. A targeted-learning roadmap for rare disease trials is given in Section 4. Section 5 presents some case studies of using RWD and RWE to support regulatory decisions in various settings. A discussion and some concluding remarks are given in Section 6.

## 2 Regulatory Guidance and Related Documents on Use of RWD and RWE

In the past few years, regulatory agencies have initiated numerous activities and programs and released guiding documents on the use of RWD and RWE, many of which are for general product development and regulatory decision-making. However, the principles and ideas in these programs and guidance can also be tailored into rare disease drug development. This section provides a summary of these initiatives and guidance issued by major regulatory agencies (including the International Council for Harmonisation of Technical Requirements for Pharmaceuticals for Human Use or ICH) with a special focus on how RWD and RWE can be used in rare disease drug development.

### 2.1 Guidance and related documents by the ICH

The ICH issued in March 2022 the final concept paper on general principles on plan, design, and analysis of pharmacoepidemiological studies that utilize real-world data for safety assessment of medicines [15]. This concept paper is intended to provide an internationally harmonised approach on general considerations and recommendations for use of RWD for drug, vaccine and other biologic product safety assessments. The ICH published in September 2023 a reflection paper outlining a strategic approach to address some challenges faced by regulatory agencies across the globe, such as lack of standardization of RWD/RWE terminology and formats, the heterogeneity of data quality of across RWD sources, and the various study designs used on different types of diseases, medicines, and regulatory context [16]. The reflection paper aims to (1) engage the ICH on convergence of terminology for RWD and RWE on the format for protocols and reports of study results submitted to regulatory agencies and (2) to inform the assessment of RWD and RWE for regulatory purposes.

### 2.2 Guidance and related documents by the FDA

The Framework for FDA's RWE Program [2] provides the scope of current use of RWD for RWE generation, outlines the framework for evaluating RWD and RWE for use in regulatory decision-making, and presents some of the FDA's demonstration projects on this regard. Under the RWE framework, the FDA conducts numerous RWD and RWE-focused demonstration projects through multiple venues and programs aiming to promote shared learning and understanding with external stakeholders in this blooming field [17]. In addition, the regulatory agency started since October 2022 an Advancing RWE Program which provides sponsors with the opportunity to meet with agency staff to discuss the use of RWE in product development [18]. Specifically, the Advancing RWE Program is designed to (1) identify approaches for generating RWE that meet regulatory requirements in support of labeling for effectiveness or for meeting post-approval study requirements, (2) develop agency processes that promote consistent decision-making and shared learning, and (3) promote awareness of characteristics of RWE that can support regulatory decisions. Under the FDA's Advancing RWE Program, the Oncology Center of Excellence (OCE) initiated an Oncology RWE Program aiming at fostering regulatory science and collaboration to translation RWD into RWE in oncology product development [19]. The strategic priorities of the OCE RWE Program are (1) to optimize knowledge building through a centralized oncology RWD research portfolio to ensures study efficiency, transparency, and diversity, (2) to advance the scientific development of resources, regulatory policy, and guidance on appropriate use of RWD informed by methodological research and collaboration, (3) to collaborate through strategic partnerships that foster pragmatic, appropriate use of RWD, and (4) to accelerate the field of Oncology RWE through leadership and training in rigorous evaluation, methods development, and regulatory science.

Built upon the framework of RWE program, the FDA has published the following guidance relating to RWD/RWE:

- *Use of Electronic Health Record Data in Clinical Investigations* [20]. This draft guidance first defines the realm of electronic health records (EHRs) and elucidates information contained in an EHR (e.g., a patients medical history, diagnoses, treatment plans, immunization dates, allergies, radiology images, pharmacy records, and laboratory and test results). The purpose of this guidance is intended to provide recommendations on (1) whether and how EHRs can be used as a data source in prospective clinical investigation, (2) inter-operability of using electronic data capture (EDC) in EHR systems for clinical investigation, and (3) the quality and integrity of EHR data that meet FDA's inspection, record keeping and retention requirements. This guid-ance can be useful in identifying rare disease patients, studying the natural history of a rare disease, or selecting external control patients.

- *Real-World Data: Assessing Electronic Health Records and Medical Claims Data To Support Regulatory Decision-Making for Drug and Biological Products* [21]. This guidance describes considerations when using EHRs and/or medical claims data in clinical studies to support regulatory decision on product effectiveness and safety. Specifically, the guidance discusses (1) the selection of data sources that can appropriately address the study question and that can sufficiently characterize the study population, treatments of interest, endpoints measuring treatment effect, and key covariates, and (2) data provenance and quality during data accrual and data curation and into the final analysis datasets. This guidance can help assess fit-for-use EHRs and medical claims data for rare disease drug development.

- *Real-World Data: Assessing Registries to Support Regulatory Decision-Making for Drug and Biological Products* [22]. This guidance discusses considerations (1) when assessing fit-for-use registry data in regulatory decision-making by focusing on attributes of a registry for collection of relevant and reliable data, (2) when linking to other data sources (e.g., EHRs, medical claims data, digital health technologies, or other registries), and (3) for supporting FDA review of submission that include registry data. This guidance can be very useful in assessing fit-for-use registries to identify patients, to study the natural history of a rare disease, and to select external control patients in rare disease drug development.

- *Considerations for the Use of Real-World Data and Real-World Evidence to Support Regulatory Decision-Making for Drug and Biological Products* [23]. This guidance (1) discusses the applicability of the FDA's investigational new drug application (IND) regulation to various non-interventional study designs that incorporate RWD, and (2) clarifies the agency's expectation concerning studies using RWD submitted to the FDA in support of regulatory decision. This guidance can serve as a general reference for assessing applicability of IND using RWD in rare disease trial designs.

- *Submitting Documents Using Real-World Data and Real-World Evidence to FDA for Drug and Biological Products* [24]. This guidance focuses on submissions to the FDA that rely on RWD/RWE to support a regulatory decision regarding product effectiveness and/or safety, in which the proposed use of RWD/RWE are listed, such as (1) support safety and effectiveness for a product not previously approved by the FDA, (2) support labeling changes for an FDA-approved product (e.g., adding or modifying an indication, changing the dose, dose regimen, or route of administration, expanding the labeled indication to a new population, adding comparative effectiveness or safety information, etc.), and (3) help support or satisfy a post-marketing requirement (PMR) or post-marketing commitment (PMC). The guidance also gives examples of study designs that use RWD/RWE: (1) randomized controlled trials (RCTs) that use RWD to capture clinical outcomes related to effectiveness or safety, (2) single-arm trials (SATs) that use RWD in external control arm, (3) observational studies (e.g., observational cohort studies) that generate RWE to support efficacy supplement, and (4) clinical trials or observational studies that use RWD or RWE to fulfill a PMR or PMC. This guidance can be very helpful in guiding the preparation of submission documents that contain RWD/RWE for regulatory decision-making as RWD/RWE is often used in rare disease drug development as will be seen in Section 3.

- *Considerations for the Design and Conduct of Externally Controlled Trials for Drug and Biological Products* [25]. This draft guidance provides considerations for the design and analysis of clinical trials using external controls to study the effectiveness and/or safety of drugs with discussions on (1) potential bias that threats the validity of study results, (2) the use of patient-level data from other clinical trials or from RWD sources (e.g., registries, EHRs, and medical claims), and (3) communication with and accessibility of external control data by the FDA. This guidance is particularly helpful in rare disease drug development in which trial designs often use external controls.

In addition, the FDA has also issued guidance relating use of RWE to support regulatory decision-making for medical devices [26] and data standards for drug and biological product submission containing RWD [27]. Principles and recommendations presented in these guiding documents are also applicable to rare disease drug development.

## 2.3 Guidance and related documents by the EMA

The European Medicines Agency (EMA) first published in 2018 its draft strategic goals through 2025 [28] and then formalized it in 2020 [29]. The strategic reflection paper describes five strategic goals for regulatory science of human medicines, among which is advancing patient-centered access to medicines with promoting use of high-quality RWD in decision-making as one of core recommendations. In 2019, the EMA and Heads of Medicines Agencies (HMA) set up a joint task force to describe the big data landscape and to identify practical steps for making the best use of big data in support of innovation and regulatory decision-making [30]. Under the Big Data initiative, The EMA produces a list of metadata describing RWD sources and studies to help identify and use such data in product development [31]. In addition, the EMA, together with the The European Medicines Regulatory Network, established in 2021 the Data Analysis and Real World Interrogation Network (DARWIN EU®) to support regulatory decision-making by (1) developing a catalog of observational data sources for use in medicines regulation, (2) providing a source of high-quality, validated RWD on the uses, safety and efficacy of medicines, and (3) addressing specific questions by carrying out high-quality, non-interventional studies, including developing scientific protocols, interrogating relevant data sources and interpreting and reporting study results [32]. While the EMA has published several guidelines on pharmacoepidemiology and pharmacovigilance [33, 34], the following two guiding documents issued by the EMA may be particularly useful for generating efficacy evidence in rare disease drug development:

- *Guideline on registry-based studies* [35]. The guideline provides considerations and recommendations on key methodological aspects that are specific to the use of patient registries in product marketing authorization. In particular, the guideline discusses, among others, the use of registries for evidence generation, study planning, protocol development, definition of population, data quality and analysis, and enabling and leveraging research and innovation in regulatory science for rare diseases.

- *Scientific guidance on post-authorisation efficacy studies* [36]. This guidance presents general methodological considerations (including randomized and non-randomized studies and data sources), scientific considerations for specific scenarios, and conduct post-authorization efficacy studies. In particular, the guideline discusses the potential value of using EHRs to facilitate the conduct of clinical trials to study rare disease outcomes.

## 2.4 Guidance and related documents by the NMPA

In the past few years, the China National Medical Products Administration (NMPA) has constructed real-world study (RWS) related guidance, reflecting the contribution of the NMPA to the field of RWS in drug clinical development. The following summarizes guidance documents issued by the NMPA relating to the use of RWD and RWE in drug development:

- *Guideline on Using Real World Evidence to Support Drug Development and Regulatory Decision-making* [37]. This guideline (1) defines real-world studies (RWS) related terminologies, (2) specifies the position and scope of RWE in drug development and regulatory decision-making, (3) Proposes the main sources, fit-for-purpose evaluation and curation of RWD in principle, (4) points out the main design types and research paths of RWS, (5) provides the principles of RWE evaluation in regulatory context, and (6) introduces causal inference methods commonly used in RWS. Of special mention is that the guidance states that RWD/RWE may be used to provide robust and reliable external control cohorts in single-arm trials for rare disease drug development.

- *Guideline on Real-World Study Design and Protocol Framework* [38]. This guideline (1) discusses main types of study designs for RWS including observational studies, pragmatic clinical trials, and single-arm studies, (2) describes the framework of RWS protocols, and (3) provides other considerations for RWS designs including feasibility of RWS, representativeness of the population, hybrid study design, method of virtual controls, estimands and target trial emulation. In particular, the guideline provides detailed description on the design of single-arm trials including the setup of both experimental and external control arms for rare disease drug development and regulatory decision-making.

- *Statistical Guideline on Clinical Trials for Rare Diseases* [39]. This guideline describes statistical considerations for trial designs for rare diseases, including single-arm trials and real-world studies.

- *Guideline on Using Single-arm Trials to Support Marketing Applications for Anti-tumor Drugs* [40]. This guideline discusses limitations and suitability of single-arm trials and requirements for conditional regulatory approval of anticancer drugs. The principle and approaches presented in the guideline are also applicable to rare disease drug development.

- *Guidance on Natural History Studies of Rare Diseases for Drug Development* [41].

See also Wang et al. [42] for more general description on RWE-related guidance documents by the NMPA.

## 2.5 Guidance and related documents by the Ministry of Health, Labour, and Welfare of Japan

The Ministry of Health, Labour and Welfare (MHLW) of Japan released in 2021 "Basic Principles on Utilization of Registry for Applications" [43] which highlights the main use of registry data in the following five areas: (1) planning a clinical study, (2) serving as an external control, (3) complementing or substituting a clinical study, (4) supporting evaluation of medical products for conditional approval, and (5) evaluating efficacy and safety in post-marketing studies. The Basic Principles specifically acknowledges the utility of registry data on rare diseases in each of the above applications and provides some general points-to-consider when using registry data including protection of personal information and patient consent, reliability, (appropriateness, and early consultation and specific points-to-consider when using registry data as an external control such as appropriateness of patient population, endpoints, evaluation period, statistical methods, and type of observational study for natural history. In addition, the MHLW released points to consider for ensuring the reliability in utilization of registry data [44], questions and answers on points to consider for ensuring reliability of registry data for drugs [45] and for regenerative medical products [46].

# 3 Current Practice on Use of RWD and RWE in Rare Diseases Drug Development

Whereas approvals of treatments for rare disease must meet the same statutory standards for efficacy and safety, the FDA exercises flexibility within these limits and considering treatment for a serious disease with no available therapy, or for a rare disease with limited sample size. The FDA also considers utilizing RWD/RWE, historical controls, and/or natural history studies in rare disease drug development. This section describes the current practice of using RWD and RWE along the clinical development spectrum from natural history studies and the design, conduct, and analysis of clinical trials in rare disease drug development.

## 3.1 Natural history studies

Fully understanding of the natural history of a disease is essential in drug development for the target disease, e.g., assisting trial design, precisely identifying patient population, and determining appropriate endpoints for assessment of treatment effects. For many rare diseases, however, the natural histories remain relatively unknown [47, 48], which hampers drug development for these rare diseases [12]. The FDA guidance on natural history studies for rare diseases provides a detailed description on the use, type, study protocol, data elements, and research plan of a natural history study (NHS) for rare diseases [49].

A natural history study (NHS) is "a preplanned observational study intended to track the course of the disease [from the onset until either the disease's resolution or the individual death]...to identify demographic, genetic, environmental, and other variables (e.g., treatment modalities, concomitant medications) that correlate with the disease's development and outcome" [49], in which patients may receive the current standard of care and/or emergent care. Among the broad scope in the natural history of a rare disease, disease progression, rather than onset, is perhaps most important for clinical trials. This defines the goals of an NHS for rare disease drug development as (1) identification of various disease manifestations or outcomes and pathways leading to these outcomes and responsive to an intervention and (2) understanding the mechanism of action for a drug that will alter the disease progression [50]. An NHS can be designed either retrospectively using existing data (e.g., disease registries, EHRs/EMRs) or prospectively with pre-specified study protocol to collect desired data elements. From data collection perspective, an NHS can be either cross-sectional (i.e., data are collected during a specified, limited time period) or longitudinal (i.e., data are collected at several time points over time), each of which has its own pros and cons and may be suitable to specific objectives. Although relatively more efficient with less turn-around timeframe as compared with longitudinal design, a cross-sectional NHS may suffer from sampling bias (e.g., subjects who takes a medicine treating a rare disease may not be in the study if the clinical manifestations of the disease disappear) and the inability to characterize rapidly progressive rare diseases [49, 50].

Unlike in clinical trials where causal estimands are generally constructed, an NHS often uses descriptive estimands as determined by the study objectives which may include, e.g., (1) identify demographic, genetic, environmental, and other factors that correlate with disease development and outcomes, (2) better characterize the disease and patient population, (3) clarify the impact of the disease on the lives of patients and their families, or (4) collect clinical outcomes including patient-reported outcomes that are specific for the disease. The construction of an NHS estimand can be carried out through precisely defining the following essential elements: (1) population: precision diagnostic tools (e.g., clinical manifestations, genome sequencing) are necessary to determine the study population with the disease of different subtypes at various stages, (2) treatment: any treatments that are used by the individuals should be clearly recorded for detailed treatment regimens (e.g., timing, dosage, duration, etc.), (3) endpoints: a broad scope of variables (e.g., surrogate, intermediate, and clinical) may be collected, depending on the study objectives, to support clinical studies and regulatory decision-making, and (4) population-level summary: point and distributional summaries are often used to estimate the estimands. Unlike in interventional trials in which intercurrent events (ICEs) play an important role in comparative analyses of treatment effects as they may impact the existence and/or measurement of endpoints, ICEs play minimal roles in an NHS because any ICEs observed after an individual enters into the NHS can be considered as part of the natural history or routine factors that affect the natural history of the disease.

In addition to descriptive estimands, the protocol should also specify (1) criteria and tools used for disease diagnosis, (2) detailed information (including time-independent and time-dependent covariates, standard of care received, biological test results, disease manifestations and other endpoints, etc.) to be collected, (3) methods for data collection and storage, and (4) analytical plan that addresses the following points: descriptive objectives, analysis population, definition of endpoints, testable hypothesis, and statistical methods to be used. Note that the natural history may be affected by a competing risk/event such as death, which could be quite common in follow-up of rare disease cohorts. In general, competing events as a particular type of ICEs differ from other censoring events, as the former prevents the study endpoints from occurring while the latter prevents the endpoints from being measured [51]. In this sense, treating a competing event as another form of censoring in an NHS may be inappropriate because a competing event may not be independent of the endpoint (time-to-event) of interest which may be part of the natural history if the competing event does not occur. Therefore, it is critical at the design stage of an NHS to (1) clearly define the endpoints measuring the natural history of the rare disease, (2) identify all possible ICEs that could occur during follow-up, (3) determine whether there exist competing events (e.g., death, organ transplant, etc.), and (4) specify handling strategies for different type of ICEs, e.g., a composite strategy to handle competing events if they are part of endpoint measures or a hypothetical strategy to address competing events in a world as they would not occur. See also Jewell [50] for more discussion on handling of competing risks in NHS.

## 3.2 Trial design

FDA [25] provides general discussions on the design and data considerations for externally controlled trials and Chen et al. [10] discuss the rationales of using RWD and RWE in the design and analysis of clinical trials and considerations for study planning in particular. For rare disease clinical trials, the use of RWD and RWE can be roughly classified into three areas: informing trial design, serving as external controls, and augmenting the internal control of an RCT. The rest of this section provides further considerations on each of these three areas.

*Informing trial designs.* Chen et al. [10] summarize the use of RWD and RWE to inform clinical trial design in two aspects, i.e., logistic aspects (e.g., site selection, trial monitoring, case report form design, subject enrollment planning, etc.) and scientific aspects (e.g., defining study patient population, endpoint selection, biomarker development and validation, hypothesis generation, sample size calculation, etc.), which are also applicable to rare disease drug development. For example, (1) NLRP3 (NOD-containing protein-like receptors) mutations associated with a group of rare hereditary auto-inflammatory diseases that can be identified through NHS can help precisely define the target population with different disease phenotypes [52], (2) studies of Fabry disease registry (https://www.rethinkfabry.com) and a registry of the Vasculitis Patient-Powered Research Network (VPPRN) (https://www.vasculitisfoundation.org/research/vpprn/) provide combining information for modeling efficacy endpoints in untreated or placebo subjects, which helps clinical trials for endpoint selection [53], and (3) an NHS of amyotrophic lateral sclerosis (ALS) using retrospective RWD among patients with superoxide dismutase 1 gene (SOD1) mutation suggests a median survival of 1.2 years, which helps determine the sample size and follow-up time for clinical trials in ALS disease [54]. See also Dagenais et al. [55] for more discussion and examples on the use of RWD/RWE in clinical trial design in general.

*Single-arm trials with external controls.* Single-arm trials (SATs) using RWD/RWE as external controls are perhaps one of the most important trial designs for accelerated drug development programs including rare disease drug development due to the challenges discussed in Chen et al. [12]. Depending on temporality of the control data generation relative to the start of the SAT, external control data can be retrospective (historically collected data), prospective (prospectively collected data), or retro-prospective (including both historically and prospectively collected data). Chen et al. [56] classify external controls into seven different types, each of which has its unique features, applicability, merits, and limitations. The choice of an appropriate external control type will depend on the study objectives and design, disease areas, the landscape of SoC, and availability of RWD/RWE. For example, (1) a historical control can be chosen if the landscape of clinical practice (e.g., therapeutic landscape including SoC, disease diagnostic tool and criteria) does not change rapidly; (2) a synthetic control may be considered if multiple external data sources are available in which a synthetic method can be used to create a *counterfactual* control group; and (3) a virtual control may be helpful if there exists a huge unmet medical need and a group of treatment-naive patients is available from which a prediction model can be constructed and then used to generate the *counterfactual outcomes* by plugging the data of treated patients in the single-arm study [57]. However, cautions should be taken when designing an SAT with external controls due to its inherent limitations. See Lu et al. [58] for appropriateness and conditions of single-arm trials with external controls in support of drug development and regulatory decision for rare diseases.

*A hybrid control for RCTs*. RWD/RWE can be combined with internal concurrent control data of an RCT to form a *hybrid control* (or internal control augmentation) using a pre-defined algorithms, e.g., test-and-pool [59] and Bayesian borrowing [60]. The idea of hybrid control originates from Stuart and Rubin [61] who propose an algorithm to obtain matches from multiple control groups using propensity score (PS) approach. A hybrid control design can be particularly useful for RCTs in rare diseases drug development when there is preliminary evidence showing a favorable benefit-risk profile of an IP, in which case an RCT can be designed with a larger treatment group and a smaller control group, with the latter being augmented by external control data. The obvious advantages of a hybrid control design are that (1) it allows more patients to receive a promising therapy and (2) it can provide the opportunity to assess similarity between internal control patients and external control patients and then decide the amount of external information to be borrowed. For general discussions on the use of hybrid control in precision medicine (including orphan drugs) development, see Chen et al. [56], Mishra-Kalyani et al. [62] and Freidlin and Korn [63].

## 3.3  Trial conduct

Chen et al. [64] discuss the use of RWD and RWE to improve the conduct of oncology clinical trials including patient identification, accrual, and retention, which is also applicable to rare disease trials. For example,

- *Patient engagement and trial promotion.* Leveraging large RWD datasets can help identify and engage patients with a rare disease for trial participation and promotion, e.g., linking target patients to available clinical trials with pre-prepared advertisement materials through an organized effort of a healthcare system [65]. This may help improve diversity of participation as some patients (e.g., those who are unaware of the clinical trial) may not otherwise have a chance to participate.

- *Trial planning.* RWD (e.g., patient registries for rare cancers, EHRs/EMRs) can

be used to provide the distribution of rare disease patients and healthcare providers across different geographic locations (e.g., by state) and demographic factors (e.g., age, gender, ethnicity), which can help set eligibility criteria, choose trial sites, select investigators who have experience in a particular disease area, plan patient enrollment, and set target recruitment efforts [55, 66].

- *Patient identification.* Some RWD sources (e.g., disease registries, EHRs/EMRs) may contain multiomics information (e.g., genomics, proteomics, metaboliomics, etc.), pharmacokinetics and pharmacodynamics (PK/PD) results, and/or observed toxicities, which can be used to help precisely locate or identify patients who would most likely benefit from a treatment, especially a targeted therapy (e.g., rare cancer patients carrying certain genomic features), or experience a harm if a particular treatment or its high dose is given to patients with certain characteristics [67, 68]. In addition, multiomics data can greatly help increase the number of patients gaining access to clinical trials of rare diseases with shared molecular drug target, e.g., basket trials targeting shared molecular etiologies [69, 70].

## 3.4 Analytical methods for data analysis

Analytical methods for data analysis depends on the study objectives, corresponding estimands, associated study design, observed data distribution, and estimators to be used to estimate the estimands. The methods used for rare disease drug development can roughly be classified into two categories: (1) methods used to answer descriptive questions (with descriptive estimands) and (2) methods used to answer comparative questions (with causal estimands). The main purpose for category (1) methods is to describe the status of a rare disease including prevalence and/or incidence, patient characteristics, clinical manifestation (outcomes), biomarkers (surrogates) for disease diagnosis, and potential risk factors (which could be biomarkers) associated with disease progression. Therefore, category (1) methods are often descriptive, e.g., Brookmeyer [71], Jewell [50] and references therein, with diagnostic accuracy, estimate precision, patient representativeness, and understanding of disease pathophysiology as the primary focuses. On the contrary, the main purpose for category (2) methods is to provide comparative effectiveness of a product of interest relative to another product (often an SoC) in a (hybrid) clinical trial setting (e.g., an externally controlled trial) or completely real-world non-interventional setting. Hence, category (2) methods are often comparative with causal biases as primary concerns.

There is rich literature describing the methods for estimation of causal effectiveness of medical products, e.g., Chen et al. [10], Ho et al. [72], Levenson et al. [73], He et al. [74] and references therein. The choice of analytical methods for causal effectiveness estimation depends not only on the assumptions necessary for causal estimation based on the observed data distributions and causal gaps (the difference between the true value of statistical and causal estimands [75]), but also on the study design. Clinical studies for rare disease drug development often use one or a combination of the following three study designs: (a) an RCT, with or without internal control augmentation, (b) a SAT with an external control, and (c) an observational (non-interventional) study, either retrospective or prospective. Analytical methods commonly used in rare disease clinical studies for causal effect estimation may encompass, e.g., the test-then-pool method [59] in RCT setting, Bayesian hierarchical modeling [76], Bayesian borrowing with power priors [77] or commensurate priors [78], regression adjustment (e.g., the G-computing method of Robins [79]), matching and stratification [80, 81], inverse probability (of treatment or censoring) weighting [82, 83], and targeted maximum likelihood estimation van der Laan [84]. Some data-adaptive approaches such as classification and regression tree (CART) and Bayesian additive regression tree (BART), and random forests may also be considered for causal effectiveness estimation for rare disease drugs. Steps 2 and 3 in the next section provide more discussion on analytical methods and estimation of causal estimands.

## 4 Targeted Learning Roadmap for Rare Disease Trials

Careful study design and attention to data quality can mitigate difficulties in producing reliable RWE in the rare disease setting. The Targeted Learning (TL) estimation roadmap offers a step-by-step guide to learning from data that combines principles of causal inference and statistical theory [13, 14, 72, 85]. Following the roadmap documents the entire chain of reasoning from scientific question through study design, data acquisition, statistical estimation, sensitivity analysis, and level of support for a causal interpretation of the study finding. This elevates the notion of RWE to include not only a study finding, but also the rich context for understanding how well the study met its intended goals.

## 4.1 Step 0: Specifying the research question and corresponding estimands

Learning from data begins with a well-formed scientific question. Step 0 in the TL Roadmap is largely the realm of study sponsors and regulators who design the study and the plan for acquiring high quality data [21, 86–88]. Target trial emulation [89, 90] is a useful paradigm for this prerequisite step, which culminates in a clear statement of the study question, precise specifications of the study and target populations, treatment, comparator, intercurrent events, and outcome consistent with ICH-E9 (R1) guidelines for defining an estimand [91], and a description of the data. Decisions made in this step can either mitigate or exacerbate challenges to obtaining interpretable RWE from RWD. For example, *when little is known about the natural history of a rare disease*, RWD from registries or online communities could inform study inclusion/exclusion criteria that clearly defines the study population. Enrolling a study population that is representative of the target population of interest (presumably all patients with the rare disease) enhances generalizability, but can be challenging when the source patient pool is small [92]. The slow, progressive nature of many rare diseases suggests enrollment efforts might be more successful if studies address one or more outcomes that are meaningful to patients, or use of a surrogate endpoint to avoid lengthy follow-up. For long-term outcomes where a RCT is not feasible a pragmatic trial using standardized endpoint definitions that can be ascertained accurately across a variety of RWD sources [6, 93]

## 4.2 Step 1: Defining the observed data distribution and corresponding statistical model

Step 1 of the roadmap is defining a realistic statistical model of the data generating distribution based on the time ordering of the data. A statistical model is a collection of possible probability distributions of the data. In the absence of external knowledge, the statistical model contains all joint distributions of the data. For example, given observations $O = (W, A, Y)$, with baseline covariates $W$, treatment indicator $A$, and outcome $Y$, the data likelihood can be factorized as $P(O) = P(Y|A, W)P(A|W)P(W)$. However, knowledge allows us to restrict the statistical model to distributions consistent with reality. For example, if there is a known interaction between the study drug and another medica-tion, any parametric model for $P(Y|A, W)$ that omits the drug-drug interaction term can be excluded. The statistical model can be further refined based on understanding developed in the next step of the roadmap.

## 4.3 Step 2: Mapping causal relationship of variables in the statistical model using domain knowledge

Step 2 harnesses domain knowledge to craft a causal model illustrating potential causal relationships among all variables. The structure of a causal directed acyclic graph (DAG) depicting variables as nodes and directed edges as causal relationships reveals potential confounders of the treatment-outcome association [94]. Knowledge is represented by the absence of an arrow, indicating the lack of a causal association. All arrows are present except those precluded by the time ordering or external proof. Domain knowledge is enhanced by carefully considering the process that gives rise to the data to identify potential confounders of the treatment-outcome association, issues with data completeness, potential intercurrent events, and loss to follow-up. Arrows between these nodes, treatment, and outcome depict potential sources of bias. Examining the DAG can motivate practical strategies for breaking the associations or weakening their strength. For example, the ICH E9(R1) guideline recommends considering whether it is appropriate to include an intercurrent event as part of a composite outcome, which would be reflected on a revised DAG [91].

In the rare disease setting a single arm trial must rely on an external control arm from a prior RCT, registry, other RWD source, or natural history [95]. Differences in data availability, monitoring schedules, laboratory tests, background standard of care, improved diagnostic detection of co-morbidities, and changes in the background prevalence of the outcome over time are some of the time-dependent phenomena that can confound the treatment-outcome relationship. Including these phenomena and time in the causal DAG is crucial for understanding how to answer questions about a specific intervention on the underlying system.

A causal parameter is expressed in terms of a causal contrast of counterfactual outcomes in the "full data" consistent with the causal model. If one treatment arm better represents the target population of interest than another, plan to evaluate the most relevant causal parameter: the average treatment effect among the treated (ATT), among the comparators (ATC), or among the study population (ATE). Let $Y^1$ be the outcome under exposure to the study drug and $Y^0$ be the outcome under exposure to the comparator. The ATE is given by $\psi_{\text{ATE}}^{\text{causal}} = E(Y^1) - E(Y^0)$, the ATT is given by $\psi_{\text{ATT}}^{\text{causal}} = E(Y^1|A=1) - E(Y^0|A=1)$, and the ATC is given by $\psi_{\text{ATC}}^{\text{causal}} = E(Y^1|A=0) - E(Y^0|A=0)$. In contrast to a causal parameter defined as a coefficient in a pre-specified parametric regression model, one that is defined in terms of the full data has a clear interpretation.

### 4.4 Step 3: Estimating the targeted parameter in the statistical model

Step 3 concerns estimation of a statistical parameter from observed data, in which both counterfactual outcomes are never available. Consider the ATE. The statistical parameter $\psi_{\text{ATE}}^{\text{stat},1} = E_W[E(Y|A=1) - E(Y|A=0)]$ is asymptotically equivalent to $\psi_{\text{ATE}}^{\text{causal}}$ in an ideal RCT where patients were successfully randomized, complied perfectly with their assigned treatment, and outcomes were measured accurately with no loss to follow-up. In the absence of baseline randomization, unbiased estimation requires adjusting for drivers of treatment that also affect the outcome ($W$), and $\psi_{\text{ATE}}^{\text{stat},2} = E_W[E(Y|A=1,W) - E(Y|A=0,W)]$ is expected to more closely correspond to $\psi_{\text{ATE}}^{\text{causal}}$. The causal identifying assumptions under which $\psi_{\text{ATE}}^{\text{stat},2}$ is asymptotically equivalent to $\psi_{\text{ATE}}^{\text{causal}}$ typically include consistency (the observed outcome for subject $i$ equals the counterfactual outcome under the observed treatment, $Y_i^a = Y_i|A_i = a$)), positivity (within strata defined by confounders there is a positive probability of receiving treatment at each level of interest, $0 < P(A=a|W) < 1$), and coarsening at random (CAR, "no unmeasured confounding", $Y^a \perp A|W$) [96,97].

A poorly captured outcome (e.g., unrecognized stroke, a patient-reported outcome subject to recall bias) threatens consistency. Baseline differences between the treated group and an external control group or a period effect when historical controls are used can threaten positivity. If this seems likely, one might choose to go back to step 0 and reconsider inclusion/exclusion criteria or identify an alternate data source. Or one might return to step 1 and re-define the target of estimation as the causal ATT, where the positivity assumption must hold for a more restricted population. Small sample size is often a concern in a rare disease setting, and willingness to enroll may vary with patient characteristics such as age and health status. When the external control arm is larger than the trial arm and better represents the target population, the ATC may be a better and less variable target of es-timation. Ultimately, by the end of Step 3 the statistical parameter is defined in terms of the data distribution, not with respect to a specific parametric model or estimator.

## 4.5 Step 4: Measuring uncertainty of the estimated parameter

Step 4 of the roadmap estimates $\psi^{\text{stat}}$ and provides measures of uncertainty (standard error, 95% confidence interval, $p$-value). Confidence interval coverage and nominal type-I and type-II error rates are with respect to the statistical parameter, not its causal interpretation. Targeted minimum loss-based estimation (TMLE) combined with super learning (SL) is the recommended efficient estimator [98]. Compared with propensity score-based methods, TMLE+SL reduces variance that can manifest as finite sample bias and result in wide confidence intervals. TMLE+SL also adheres to known restrictions on the statistical model, e.g., outcomes must lie within known bounds, without imposing unnecessary constraints such as homogeneity of the treatment effect. Machine learning algorithms trained on small datasets can easily overfit the data, so care must be taken when specifying the SL library and cross-validation scheme. Recommendations include using the highly adaptive lasso [99], specifying multiple candidate parametric models, avoiding algorithms known to require copious amounts of training data (e.g., neural nets), coupling algorithms with both mild and aggressive covariate pre-selection strategies, and setting the number of cross-validation folds large [100]. Simulation studies can help the analyst make informed choices.

## 4.6 Step 5: Performing sensitivity analysis

Step 5 considers how closely the effect estimate aligns with the true causal effect, and how much confidence to place in the study finding as a guide for decision-making. A non-parametric sensitivity analysis explores the impact on the effect estimate and confidence interval bounds under small and large violations of the causal assumptions linking the statistical and causal parameters. Such violations induce a causal gap between the statistical and causal parameters, a bias that will not decrease with increased sample size [101]. In some cases, an upper bound on the causal gap may be known from external sources. Otherwise, expertise is used to ascertain the plausible size and direction of the causal gap. If the confidence interval excludes the null under all plausible causal gap sizes, then this increases confidence in the substantive conclusion drawn from the study. However, when little is known about the natural history of the disease it will be difficult to determine a plausible gap.

The documentation produced by following the roadmap clearly and comprehensively connects the scientific problem to the analytic result. Each step highlights different challenges to producing reliable RWE, and promotes clear justification of strategies and assumptions underlying each decision. Examining each link in the chain of reasoning improves our ability to assess the validity of the RWE.

## 5 Case Studies

### 5.1 NHS to support product effectiveness

*Skyclarys*. Friedreich ataxia is an autosomal recessive neurodegenerative disorder causing progressive damage to the spinal cord, peripheral nerves, and the brain and resulting in ataxia, dysarthria, and areflexia of adults and children. The disease often develops in children and teenagers and gradually worsens as the patient ages. There were no approved therapies for the condition until February 28, 2023 when the FDA approved SKYCLARYS (omaveloxolone) to treat Friedreich ataxia based on a 48-week randomized, placebo-controlled, and double-blind study (NCT02255435) and an open-label extension. The study enrolled 103 individuals with Friedreich ataxia who received either placebo (52 individuals) or Skyclarys (51 individuals) to evaluate the change in the modified Friedreichs Ataxia Rating Scale (mFARS) score at week 48 [102]. Results of the study showed that subjects receiving Skyclarys performed significantly better on the mFARS scores than those receiving placebo [103]. An analysis of subjects who continued treatment with Skycalrys in the open-label extension for up to 3 years suggests a better performance on the mFARS compared to a propensity-matched set of untreated subjects from a natural history study [102, 104].

### 5.2 Use of RWD/RWE to support regulatory decision in various settings

He et al. [11] provide a comprehensive review of six case studies that focus on rare diseases with significant unmet medial needs. The use of RWD/RWE in these studies varies across different settings:

- As part of the original marketing application (Avelumab, Tafasitamab)

- As primary data source for label expansion (Prograf, SurgiMend)

- As one of the data sources for label expansion (Orencia)

- As supplemental information for the regulatory decision (Ibrance)

The rest of this section highlights the key comments by the FDA during their review and analysis by He et al. [11] of the case studies. It is crucial to emphasize that the outcomes of these case studies are intricately linked to several key factors, including adherence to regulatory guidance, utilization of fit-for-purpose data, and the incorporation of causal thinking and framework into study design and analysis. These factors have been extensively discussed and highlighted throughout this paper, underscoring their significance in determining the success or failure of the clinical studies in generating reliable and robust RWE for regulatory decision-making.

The *Avelumab* case study revealed limitations of retrospective data, specifically regarding the small sample size and selection bias in the historical control. The data were considered exploratory and only used to further understand the natural history of the disease in the target population. Consequently, the RWD did not play a pivotal role in the marketing application and approval process.

In the *Tafasitamab* case study, the key concerns were focused on data quality, completeness, and population comparability. The reliability and relevancy of the data were assessed, highlighting the need for improvements in these areas. Overall, though, the real-world studies were judged to have many limitations and only provided contextual evidence in the original FDA approval.

The *Prograf* and *Orencia* case studies made use of reputable and established registry data sources, namely SRTR and CIBMTR, respectively. The suitability of SRTR for its intended purpose was acknowledged, demonstrating that a non-interventional study, despite potential biases, can still yield robust scientific findings and meet regulatory requirements for approval. Similarly, CIBMTR, another well-known and long-running registry data source, was recognized by the FDA as fit-for-purpose RWD source. The approval of Orencia further exemplified the pivotal role of RWD/RWE in supporting regulatory decision and addressing critical unmet medical needs.

However, the *SurgiMend* case faced significant challenges as both the FDA and the Advisory Committee panel identified multiple data issues that rendered the data unsuitable for its intended purpose. These issues included the absence of crucial potential confounding variables, a substantial amount of missing data on key components of the primary endpoint, differential missingness among treatment groups, lack of validation for the composite endpoint, conflicting conclusions from sensitivity analysis, and limited information on adverse events, including death. In the *Ibrance* case, the retrospective cohort study in Flatiron Health Analytic Database was hindered by a small sample size, which prevented meaningful statistical comparisons for regulatory approval. Despite this limitation, with the favorable benefit-risk profile of Ibrance in women based on several large, randomized studies combined with the supportive RWD along with safety information from review of two phase 1 studies, and postmarketing reports by the sponsor, the FDA found a favorable benefit-risk profile for Ibrance in men.

# 6 Discussion and Concluding Remarks

The ASA BIOP RWE SWG in its phase 1 and 2 projects have provided landscape assessments on the general use of RWD and RWE in clinical development and regulatory decision-making of medical products; see Chen et al. [10]; Ho et al. [72]; Levenson et al. [73]; Levenson et al. [105]; and Chen et al. [106]. The principles and approaches presented in these papers about some key aspects, e.g., selection of fit-for-purpose RWD, prospectively defined covariates for balance score calculation, sample size determination, and type I error control, are also applied to rare disease trials. In particular, besides what discussed in Section 3, the following aspects may also be considered when using RWD and RWE in rare disease drug development:

- *Assessment of fit-for-purpose RWD.* Levenson et al. [105] presented an approach to assessing fit-for-purpose RWD with three-dimensional key elements: (1) relevancy based on a specific research question which includes disease population, response/outcome variables, treatment/exposure, confounding variables, time duration, and generalizability score, (2) reliability based on a specific research question which includes quality measures and data completeness score, and (3) fit-for-research consideration that includes data provenance/traceability and standard.

- *Construction of estimands.* The TL-based approach described in Section 4 discusses how the estimands can be precisely defined in Step 0. Some additional considerations in constructing estimands of treatment effects may include: (1) for single-arm trials with external controls, the average treatment effect among the treated (ATT) estimand may be appropriate as it helps answer the question of treatment effectiveness among those who currently receive it, (2) for approved products with intended label expansion, the average treatment effects among the untreated (ATU) estimand may be more of interest [106, 107], and (3) the complexities of ICEs (e.g., frequency, patterns) may differ substantially between the treated subjects and external control subjects, which may possibly cause biased estimates of treatment effects if not handled appropriately.

- *Possible biases.* Externally controlled trials, often used in rare disease drug development, have the disadvantage of suffering from multiple sources of biases, which need to be addressed throughout the design, conduct, analysis, and reporting of study results. Some strategies to minimize these biases, e.g., [108] in SATs, should be clearly pre-defined in the study protocol, communicated with regulatory agencies, and implemented during the study conduct and analysis.

- *Sensitivity analyses.* The last step of the TL-based roadmap is to perform sensitivity analyses to evaluate the impact of possible violations of causal assumptions on the estimated treatment effects. It is important to realize that, regardless of the study design (e.g., RCT, SAT, or hybrid control design), the sensitivity analyses should be performed in the reverse order of bias occurrence [109, 110]. For example, confounding bias in the source population (e.g., in an SAT) often occurs before selection bias (e.g., non-exchangeability) which can take place before information bias (e.g., missing data) and distributional bias (e.g., inappropriate models); in this case, the sensitivity analyses should be performed in the order of evaluating distributional assumptions, informational assumptions (e.g., missing data mechanisms), exchangeability assumptions, and no unmeasured confounding assumptions.

In summary, there is a wide range of applications of RWD and RWE in clinical development of medical products [111] and special attentions are required for their use in rare disease drug development due to the unique features of rare diseases [12]. The TL-based method provides a systematic approach to causal effect estimation in a transparent and interpretable manner and hence is recommended in RWE generation for rare disease drug development.

## Abbreviations

| | |
|---|---|
| EDC | Electronic data capture |
| EHRs | Electronic Health Records |
| EMRs | Electronic medical records |
| ICEs | Intercurrent events |
| IND | Investigational new drug |
| NHS | Natural history study |
| ODA | Orphan Drug Act |
| PD | Pharmacodynamics |
| PK | Pharmacokinetics |
| PMC | Post-marketing commitment |
| PMR | Post-marketing requirement |
| PREA | Pediatric Research Equity Act |
| RCTs | Randomized controlled trials |
| RWD | Real-world data |
| RWE | Real-world evidence |
| SATs | Single-arm trials |

## Acknowledgements

The authors thank Dr. Mark Levenson of the FDA whose constructive comments have helped improve the presentation of the paper.

## Authors' contributions

## Funding

There is no funding support to this study.

**Data availability**

No clinical data are involved in this research.

## Declaration

### Ethics approval and consent to participate

Not applicable.

### Consent for publication

Not applicable.

### Competing interest

The authors report there are no competing interests to declare. The FDA had no role in data collection, management, or analysis. The views expressed are those of the authors and not necessarily those of the US FDA.